\documentclass{article}
\usepackage{spconf,amsmath,graphicx,hyperref, amssymb}
\usepackage{bm} 
\usepackage{booktabs,multirow,}
\usepackage{threeparttable, pifont}
\usepackage{cite}
\usepackage{enumitem}
\usepackage{subcaption}
\usepackage{graphicx}

\title{Enhancing Speaker Verification with w2v-BERT 2.0 and Knowledge Distillation guided Structured Pruning}
\name{Ze Li$^{1,3}$, Ming Cheng$^{1,3}$, Ming Li$^{2,3}$$^\dagger$\thanks{$^\dagger$Corresponding Author: Ming Li. 
\\
\indent\indent 
This research is funded in part by the National Natural Science Foundation of China (62571223), Yangtze River Delta Science and Technology Innovation Community Joint Research Project (2024CSJGG1100) and Guangdong Science and Technology Plan (2023A1111120012). Many thanks for the computational resource provided by the Advanced Computing East China Sub-Center.
}}
\address{
$^{1}$School of Computer Science, Wuhan University, Wuhan, China \\
$^{2}$School of Artificial Intelligence, Wuhan University, Wuhan, China \\
$^{3}$Suzhou Municipal Key Laboratory of Multimodal Intelligent Systems, \\
Duke Kunshan University, Kunshan, China \\
\{lize389, ming.cheng\}@whu.edu.cn, ming.li369@dukekunshan.edu.cn
}
\begin{document}
\ninept
\maketitle
\begin{abstract}
Large-scale self-supervised Pre-Trained Models (PTMs) have shown significant improvements in the speaker verification (SV) task by providing rich feature representations. In this paper, we utilize w2v-BERT 2.0, a model with approximately 600 million parameters trained on 4.5 million hours of unlabeled data across 143 languages, for the SV task. The MFA structure with Layer Adapter is employed to process the multi-layer feature outputs from the PTM and extract speaker embeddings. Additionally, we incorporate LoRA for efficient fine-tuning. Our model achieves state-of-the-art results with 0.12\% and 0.55\% EER on the Vox1-O and Vox1-H test sets, respectively. Furthermore, we apply knowledge distillation guided structured pruning, reducing the model size by 80\% while achieving only a 0.04\% EER degradation. Source code and models are released at \url{https://github.com/ZXHY-82/w2v-BERT-2.0_SV}.
\end{abstract}
\begin{keywords}
speaker verificaition, w2v-BERT 2.0, LoRA, knowledge distillation, structured pruning
\end{keywords}
\section{Introduction}
\label{sec:intro}
Speaker verification (SV) aims to authenticate the identity of a speaker by analyzing the voice signal. In recent years, significant advancements in deep learning, coupled with large-scale labeled datasets~\cite{voxceleb1, vox2dev, voxblink2, cnceleb1, cnceleb2}, have led to substantial improvements in the performance of deep neural network-based SV systems~\cite{ecapa-tdnn, cam++, redimnet}. 

However, the scale of existing labeled datasets remains insufficient to meet the increasing complexity of model architectures. As a result, researchers have turned to large-scale Pre-Trained Models (PTMs)~\cite{hubert, wav2vec2.0, unispeech, w2v-bert-2.0, wavlm}, which are typically trained on hundreds of thousands or even millions of hours of unlabeled speech data. These models offer powerful feature representations that can significantly enhance performance on downstream tasks.
Chen et al.~\cite{wavlm, hubert_w2v_unis_asv} employ a layer-wise weighted average of PTM's features, followed by a speaker model such as ECAPA-TDNN~\cite{ecapa-tdnn} for the SV task. Kim et al.~\cite{wavlm_lap_astp} introduces Layer-wise Attentive Pooling, which applies time-dynamic weighting to multi-layer representations, overcoming the limitation of conventional weighted summation that ignores certain layers. Peng et al.~\cite{wavlm_ca-mhfa} introduces Context-Aware Multi-Head Factorized Attentive Pooling, which incorporates contextual information and grouped queries, thereby obtaining more robust utterance-level representations. Zhao et al.~\cite{whisper_asv} and Cai et al.~\cite{cai_asr_sv} build upon the concept of MFA-Conformer~\cite{mfa}, concatenating all or part of the features from different PTM layers to capture richer and more comprehensive speaker representations.

Previous studies focused on Transformer-based self-supervised PTMs for the SV task~\cite{wavlm, hubert_w2v_unis_asv, wavlm_lap_astp, wavlm_ca-mhfa}. In contrast, w2v-BERT 2.0~\cite{w2v-bert-2.0} is a self-supervised PTM built on a Conformer-based architecture, which has been demonstrated by MFA-Conformer~\cite{mfa} to be effective for SV and superior to the Transformer-based architecture. Furthermore, w2v-BERT 2.0 adopts a training strategy that optimizes both a contrastive loss and a masked prediction loss simultaneously, and trained on 4.5 million hours of unlabeled audio covering 143 languages, leading to strong performance on audio classification tasks.

In this work, we utilize w2v-BERT 2.0 as the encoder for the SV task. Speaker embeddings are extracted using the MFA~\cite{mfa} structure, and a Layer Adapter~\cite{cai_asr_sv} module is introduced for each layer's output before concatenation, enabling better adaptation of the PTM's output to the specific task domain. Additionally, Low-Rank Adaptation(LoRA)~\cite{lora} is employed for efficient fine-tuning. To enhance the model's practicality for real-world deployment, we apply a knowledge distillation guided structured pruning technique~\cite{prune}, which prunes the PTM while minimizing performance degradation.

The primary contributions of this paper can be summarized as follows:
\begin{itemize}[left=0pt] 
\item We are the first to apply the w2v-BERT 2.0 PTM to the SV task, achieving state-of-the-art(SOTA) results of 0.12\% and 0.55\% EER on the Vox1-O and Vox-H test sets, respectively.
\item  We employ the MFA structure, combined with the Layer Adapter and LoRA modules, to efficiently adapt the model to the SV task.
\item We utilize a knowledge distillation guided structured pruning strategy, reducing the model size by 80\% with only a 0.04\% EER degradation.
\end{itemize}

\begin{figure*}
  \centering
  \includegraphics[width=0.87\linewidth]{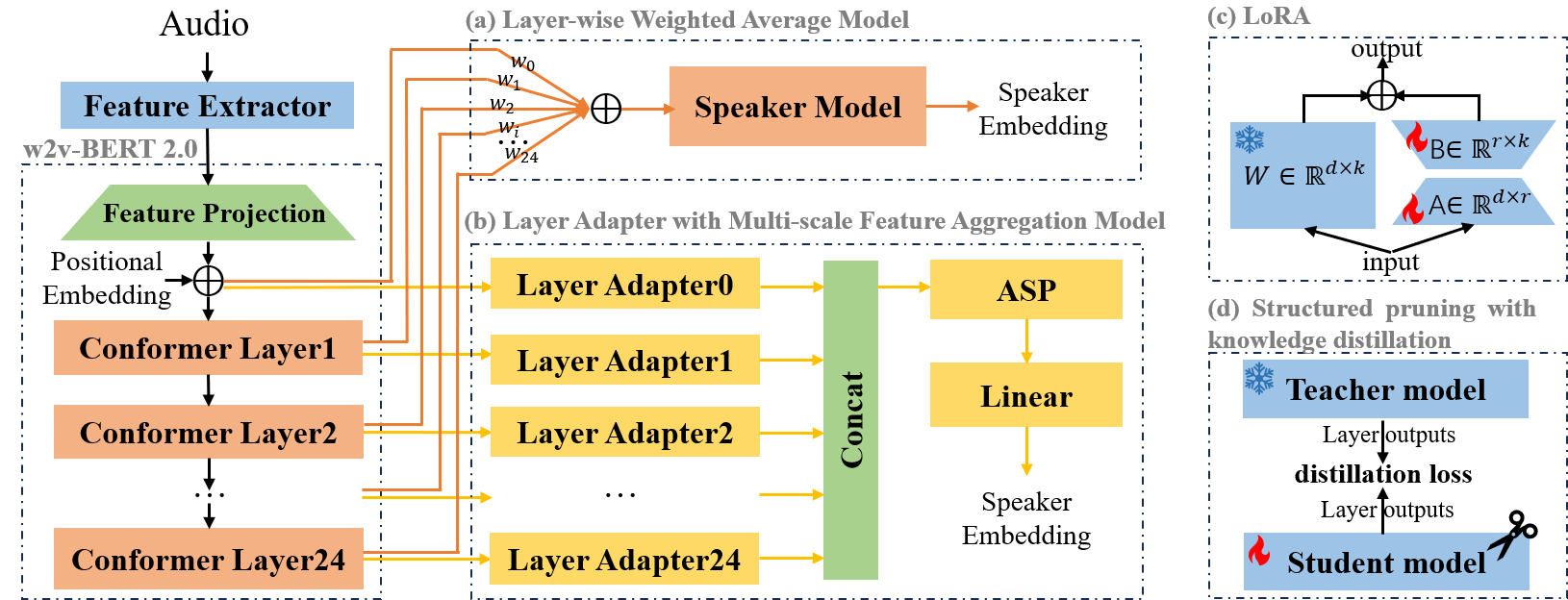}
  \caption{Module architecture for speaker verification with w2v-BERT 2.0 and knowledge distillation guided structured pruning.}
  \label{fig:system_framework}
\end{figure*}

\section{Methods}
\label{sec:method}

\subsection{Pre-trained Model: w2v-BERT 2.0}

w2v-BERT 2.0~\cite{w2v-bert-2.0} is a large-scale multilingual self-supervised model, designed for speech representation and introduced in the SeamlessM4T framework~\cite{w2v-bert-2.0}. Building on the w2v-BERT~\cite{w2v-bert} architecture, it consists of 24 conformer layers and integrates both contrastive learning and masked language modeling. The model is trained on 4.5 million hours of unlabeled audio data, covering 143 languages. In this paper, we apply w2v-BERT 2.0 for the SV task. Given a speech utterance $x$, we first extract its fbank features, and then input them into the PTM to obtain the features of each layer:
\begin{equation}
    [h_0,h_1, \ldots,h_L] = \text{W2v-BERT-2.0}(\text{Fbank($x$)})
\end{equation}
where \( h_i \in \mathbb{R}^{D \times T} \) represents the output of the \(i\)-th conformer layer, with \(D\) as the hidden dimension and \(T\) as the number of frames.

\subsection{Layer-wise Weighted Average Model}
The layer-wise weighted average approach~\cite{wavlm} is currently one of the most widely used and effective methods for fine-tuning a PTM on the SV task. In this method, each layer of the PTM is assigned a learnable weight, which is updated during training. The final frame feature $H$ is obtained by computing a weighted average of all layer outputs as shown in Fig.\ref{fig:system_framework}(a), replacing the Fbank feature fed into the speaker model for extracting the speaker embedding.
\begin{equation}
    H = \sum_{i=0}^{L} \frac{e^{w_i}}{\sum_{j=0}^{L} e^{w_j}} \cdot hi
\end{equation}
where $w_i$ is the weight of the $i$-th layer, and $h_i$ is the feature output from the $i$-th layer.

\subsection{Multi-scale Feature Aggregation Model}

Another strategy for fine-tuning PTMs is Multi-scale Feature Aggregation. Following MFA-Conformer~\cite{mfa}, the features of all layers are concatenated and fed into an Attention Statistics Pooling (ASP) module~\cite{asp_pooling}. Unlike weighted averaging, this direct concatenation preserves the full layer information, while ASP learns the relative importance across layers and dimensions. The speaker embedding $E$ is then obtained via a linear transformation of the ASP output.
\begin{equation}
\begin{aligned}
    E = \text{Linear}(\text{ASP}(\text{Concat}(h_0,h_1,\ldots,h_L)))
\end{aligned}
\end{equation}

\subsection{Layer Adapter for Model Adaptation}

Moreover, considering that directly using the raw layer features for the SV task could lead to poor generalization, we introduce a lightweight Layer Adapter~\cite{cai_asr_sv} module for each layer output before concatenation. The adapter structure consists of two linear layers followed by layer normalization and a rectified linear unit activation function. The first linear layer projects the input feature from dimension $d$ to a hidden size of $d'$, while the second linear layer maps the $d'$-dimensional representation to another $d'$-dimensional space.

\subsection{LoRA for Model Adaptation}
Compared to full fine-tuning, LoRA~\cite{lora} adapts PTMs efficiently by introducing a small number of trainable parameters in a low-rank space, reducing both computational and memory costs while maintaining effective task adaptation. In this paper, we apply LoRA to the query and value weights of PTM's self-attention module. The update mechanism for the model's weight matrix is described as follows:
\begin{equation}
    W' = W + \frac{\alpha}{r} \cdot A \cdot B
\end{equation}
where \( W \in \mathbb{R}^{d \times k} \)  is the original weight matrix, \( A \in \mathbb{R}^{d \times r} \) and \( B \in \mathbb{R}^{r \times k} \) are the low-rank matrices, $r$ is the rank of the adaptation, and  $\alpha$ is a scaling factor that controls the magnitude of the update.

\subsection{Structured Pruning with Knowledge Distillation}
The large parameter size and computational cost of PTMs pose significant challenges for deployment on resource-constrained devices. Inspired by \cite{prune}, we apply knowledge distillation guided structured pruning to the w2v-BERT 2.0 model. To preserve the original representational capacity of the model, a teacher–student framework is employed, aligning the outputs of the pruned student model with those of the unpruned teacher model. The distillation loss combines L1 and cosine distances with equal weights:
\begin{equation}
\mathcal{L}_{\text{distill}} = \sum_{l=0}^{L} \sum_{t=1}^{T} \left( L_1(h_i^t, \hat{h}_i^t) - cosine(h_i^t, \hat{h}_i^t) \right)
\end{equation}
where $L$ denotes the number of layers, $T$ denotes the number of frames, $h_i^t$ and $\hat{h}_i^t$ represent the $t$-th frame output of the $i$-th layer from the teacher and student models, respectively.

Pruning is achieved by optimizing the L0 regularization term $||\theta||_0$, where $\theta$ denotes the parameters to be pruned. However, the L0 term is discrete and non-differentiable. To address this, the parameters targeted for pruning are modeled as random variables governed by the Hard Concrete distribution, as described in \cite{prune}:
\begin{equation}
    \theta = \{ \hat{\theta_j} z_j \}_{j=1}^J, \quad z_j \sim q(z_j|\alpha_j)
\end{equation}
where $\hat{\theta_j}$ denotes the $j$-th group of prunable parameters, and $z_j$ is a stochastic binary gate sampled from the Hard Concrete distribution:
\begin{equation}
\begin{aligned}
    s_j = \sigma(\frac{\log u_j-\log (1-u_j)+\log\alpha_j}{\beta}), \\
    z_j = \min(1,\max(0,(\zeta-\gamma)\cdot s_j+\gamma))
\end{aligned}
\end{equation}
where $u_j$ is drawn from a uniform distribution, $\beta$ is a temperature parameter controlling the smoothness of $s_j$, and $\zeta$ and $\gamma$ control the upper and lower bounds of $s_j$. We set $\beta = 2/3$, $\gamma = -0.1$, and $\zeta = 1.1$.
Finally, the expected value of the L0 norm is given by:
\begin{equation}
\mathbb{E}_{q(\theta|\hat{\theta},\alpha)}[||\theta||_0] = \sum_{j=1}^{|G|}|g|\cdot \sigma(\log\alpha_j-\beta\log\frac{-\gamma}{\zeta})
\end{equation}
where $|G|$ denotes the number of groups and $|g|$ is the number of parameters in the $g$-th group.

The final loss function employs the augmented Lagrangian method~\cite{lglr} for more effective fine-grained control of the sparsity in the pruned model:
\begin{equation}
    \max_{\lambda_1,\lambda_2} \min_{\hat{\theta},\alpha} \mathbb{E}_{q(\theta|\hat{\theta,\alpha})}[\mathcal{L}_{\text{distill}} + \lambda_1(||\theta||_0-t) + \lambda_2(||\theta||_0-t)^2]
\end{equation}
where $\lambda_1$ and $\lambda_2$ are learnable Lagrange multipliers and t represents the predefined target sparsity.

\begin{table*}[htbp]\centering 
    \scriptsize 
    \caption{Performance comparison of the w2v-BERT 2.0 based SV model with other SV models.}
    \tabcolsep=0.4em
     \label{tab:res}
\begin{threeparttable}
    \begin{tabular}{lcccccccccccc}
    \toprule

    \multirow{2}*{\textbf{Frontend}} &\multirow{2}*{\textbf{Model}} &\multirow{2}*{\textbf{Params}} & \multirow{2}*{\textbf{LMFT}} & \multirow{2}*{\textbf{\parbox{1cm}{\centering Score \\ calibration}}} & \multicolumn{2}{c}{\textbf{Vox1-O}} & \multicolumn{2}{c}{\textbf{Vox1-E}} & \multicolumn{2}{c}{\textbf{Vox1-H}} & \multicolumn{2}{c}{\textbf{CN-Celeb Test}} \\
    \cmidrule(lr){6-7} \cmidrule(lr){8-9} \cmidrule(lr){10-11} \cmidrule(lr){12-13} & & & & & \textbf{EER(\%)} & \textbf{mDCF} & \textbf{EER(\%)} & \textbf{mDCF} & \textbf{EER(\%)} & \textbf{mDCF} & \textbf{EER(\%)} & \textbf{mDCF} \\
    \midrule
     \multirow{4}*{Fbank} & ECAPA-TDNN(C=1024)~\cite{ecapa-tdnn} & 14.7M & $\times$ & $\times$ & 0.87 & 0.107 & 1.12 & 0.132 & 2.12 & 0.210 & - & - \\
     & CAM++~\cite{cam++} & 7.2M & $\times$ & $\times$ & 0.73 & 0.091 & 0.89 & 0.100 & 1.76 & 0.173 & 6.78$^\ddagger$ & 0.383$^\ddagger$ \\
     & ReDimNet-B6~\cite{redimnet} & 15.0M & $\checkmark$ & $\checkmark$ & 0.37 & 0.030 & 0.53 & 0.051 & 1.00 & 0.097 & - & - \\
     & ERes2NetV2~\cite{3dspeaker} & 17.8M & $\checkmark$ & $\times$ & 0.61 & 0.054 & 0.76 & 0.082 & 1.45 & 0.143 & 6.04$^\ddagger$ & 0.362$^\ddagger$ \\
     & ResNet221~\cite{wespeaker} & 23.8M & $\checkmark$ & $\checkmark$ & 0.51 & - & 0.68 & - & 1.21 & - & 5.66$^\ddagger$ & 0.330$^\ddagger$ \\
     & ResNet293~\cite{voxblink2} & 98.9M & $\checkmark$ & $\checkmark$ & 0.17* & \textbf{0.006*} & 0.37* & 0.037* & 0.68* & 0.070* & - & - \\
    \midrule
    HuBERT Large & ECAPA-TDNN(C=512)~\cite{hubert_w2v_unis_asv} & 317+8.8M  & $\checkmark$ & $\checkmark$ & 0.59 & - & 0.65 & - & 1.23 & - & - & - \\
    \midrule
    Wav2Wec2.0 Large & ECAPA-TDNN(C=512)~\cite{hubert_w2v_unis_asv} & 317+8.8M & $\checkmark$ & $\checkmark$ & 0.59 & - & 0.63 & - & 1.14 & - & - & - \\
    \midrule
    UniSpeech-SAT Large & ECAPA-TDNN(C=512)~\cite{hubert_w2v_unis_asv} & 317+8.8M & $\checkmark$ & $\checkmark$ & 0.54 & - & 0.57 & - & 1.18 & - & - & - \\
    \midrule
    \multirow{3}*{WavLM Large} & ECAPA-TDNN(C=512)~\cite{wavlm} & 317+8.8M  & $\checkmark$ & $\checkmark$ & 0.38 & - & 0.48 & - & 0.99 & - & - & - \\
     & CA-MHFA~\cite{wavlm_ca-mhfa} & 317+2.3M & $\checkmark$ & $\checkmark$ & 0.42 & - & 0.48 & - & 0.96 & - & - & - \\
     & LAP+ASTP~\cite{wavlm_lap_astp} & 317+2.3M & $\checkmark$ & $\checkmark$ & 0.37 & 0.059 & 0.50 & 0.055 & 1.01 & 0.099 & - & - \\
    \midrule
     Nemo Large & MFA~\cite{cai_asr_sv} & 131M & $\checkmark$ & $\checkmark$ & 0.43 & 0.062 & 0.66 & 0.071 & 1.35 & 0.135 & - & - \\
    \midrule
     & \multirow{3}*{LoRA\_Adapter\_MFA} & \multirow{3}*{580+6.2M}  & $\times$ & $\times$ & 0.23* & 0.029* & 0.38* & 0.040* & 0.81* & 0.082* & \textbf{4.67}$^\ddagger$ & \textbf{0.297}$^\ddagger$ \\
     & &  & $\checkmark$ & $\times$ & 0.14* & 0.020* & 0.31* & 0.032* & 0.73* & 0.071* & - & -\\
     \multirow{-3}*{w2v-BERT 2.0} & &  & $\checkmark$ & $\checkmark$ & \textbf{0.12*} & 0.025* & \textbf{0.27*} & \textbf{0.028*} & \textbf{0.55*} & \textbf{0.051*} & - & - \\
     
    \bottomrule
    \end{tabular}
    \begin{tablenotes}
        \item * indicates results obtained using the VoxCeleb2 and VoxBlink2 datasets for training. 
        $^\ddagger$ indicates results obtained using only the CN-Celeb1\&2 datasets for training.

    \end{tablenotes}
\end{threeparttable}
\end{table*}

\section{Experiments}
\subsection{Datasets}
The experiments are conducted using the VoxCeleb1\&2~\cite{voxceleb1, vox2dev}, VoxBlink2~\cite{voxblink2} and CN-Celeb1\&2~\cite{cnceleb1,cnceleb2} datasets. For VoxCeleb model training, we utilize the VoxCeleb2 development set and the VoxBlink2 dataset. During the evaluation phase, both the VoxCeleb1 development and test sets are used. The SV performance is evaluated based on three official trial lists: Vox1-O, Vox1-E and Vox1-H. For CN-Celeb, only the development sets of CN-Celeb1 and CN-Celeb2 are used for training. We choose to average all the embeddings that belong to the same enrollment speaker to get the final speaker embedding for the CN-Celeb test set evaluation.

\subsection{Model Configuration}
\label{sec:model_config}
We use w2v-BERT 2.0 as the encoder to extract features from each layer, followed by the design of four distinct modules for speaker embedding extraction, as described below:

\textbf{Layer-wise Weighted Average Model:} Similar to \cite{wavlm}, we utilize the small ECAPA-TDNN model as the speaker model to process the weighted average of all layer outputs.

\textbf{MFA Model:} This model consists of an ASP module and a Linear module, where the outputs from all layers are concatenated and directly fed into the ASP module. The speaker embedding dimension is set to 256, and the hidden dimension of the ASP module is matched to the layer dimension.

\textbf{Layer Adapter with MFA model:} Building upon the MFA model, we add a Layer Adapter module after each layer. The hidden dimension $d'$ of the Layer Adapter is set to 128, while the ASP module’s hidden dimension is also set to $d'$.

\textbf{LoRA with Layer Adapter and MFA model:} In this model, LoRA is applied to the query and value linear layers of the self-attention modules in each Conformer layer of the PTM. The rank $r$ is set to 64, and the weight scaling factor $\alpha$ is set to 128.

\subsection{Training Details}
\label{sec:train_detail}
Our training process is divided into three stages as follows:

\textbf{i) PTM freeze training:}  
In this phase, the PTM is frozen. The acoustic features are 80-dimensional fbank coefficients with a frame length of 25ms and a hop size of 10ms. These features are normalized using mean and standard deviation before being fed into the PTM.
On-the-fly data augmentation~\cite{on-the-fly} is applied by adding background noise or convolutional reverberation noise. The MUSAN~\cite{musan} and RIR Noise~\cite{RIR} datasets are used as noise sources and room impulse response functions, respectively. The speed perturbation~\cite{speed_aug}, which speeds up or down each utterance by a factor of 0.9 or 1.1, is applied to yield shifted pitch utterances that are considered from new speakers, but is not utilized during training with the VoxBlink2 dataset. AdamW~\cite{adamw} optimizer with weight decay of 1e-4 is used, along with a StepLR scheduler with 5 epochs decay. The learning rate starts at 1e-4 and decreases to 1e-5, with a decay factor of 0.1. The margin and scale of ArcFace~\cite{arcface} are set to 0.2 and 32, respectively. A linear warm-up learning rate schedule is used for the first 5 epochs to stabilize training. The input frame length is set between 200 and 300.

\textbf{ii) Joint fine-tuning:} 
Subsequently, the PTM is unfrozen for fine-tuning. The learning rate starts at 1e-5 and decays to 5e-6 using a cosine decay schedule over 2 epochs, with a total of 4 epochs dedicated to fine-tuning.

\textbf{iii) Large margin fine-tuning and score calibration:} 
In this stage, the Large-Margin Fine-Tune (LMFT)~\cite{lmft} strategy is introduced, using only the VoxCeleb2 dataset. All data augmentation strategies are stopped. The input frame length is set between 500 and 600. For ArcFace, a margin of 0.5 is applied. The learning rate starts at 1e-5 and decays to 5e-6 using a cosine decay schedule over 1 epoch, with a total of 2 epochs dedicated to fine-tuning. Additionally, AS-norm~\cite{As-norm} and QMF~\cite{qmf} are used for scoring calibration.

\subsection{Pruning Details}
We perform structured pruning on the joint fine-tuned w2v-BERT 2.0 model, focusing on the FFN intermediate dimensions, convolution channels, and the number of attention heads of each Conformer layer. 
First, the teacher-student framework is initialized with the w2v-BERT 2.0 model, the teacher model remains frozen. The target sparsity increases linearly to the pre-set value over the first 10,000 items. The total number of epochs is 20. AdamW optimizer is used, with a learning rate of 2e-4 and 2e-2 for the student model parameters and sparsity-related parameters, respectively. After pruning, the pruned student model is further distilled by an additional 20 epochs from the teacher model. Finally, the pruned student model replaces the initial joint fine-tuned w2v-BERT 2.0 model, and further fine-tuning on the SV task based on the stages outlined in Section~\ref{sec:train_detail}.





\section{Results}

\begin{table}[t]\centering \scriptsize
    \caption{Performance comparison of different w2v-BERT 2.0 based model architectures.}
    \tabcolsep=0.4em
     \label{tab:res1}
    \begin{tabular}{lccc}
    \toprule
    \textbf{Model} & \textbf{Data} & \textbf{Params} &\textbf{Vox1-O EER}\\
    \midrule 
     ECAPA-TDNN(freeze PTM) & &  & 0.49\% \\
     + LMFT (Joint Fine-tuning) & & & 0.26\% \\
     + Joint Fine-tuning & & & 0.29\% \\
     ++ LMFT & \multirow{-4}*{VoxCeleb2} & \multirow{-4}*{580+8.8M} & 0.22\% \\
     \midrule
     MFA(freeze PTM) & &  & 0.46\%\\
     + LMFT (Joint Fine-tuning) & & & 0.28\% \\
     + Joint Fine-tuning & & & 0.38\% \\
     ++ LMFT & \multirow{-4}*{VoxCeleb2} & \multirow{-4}*{580+65.6M} & 0.26\% \\
     \midrule
     Adapter\_MFA(freeze PTM) & &  & 0.43\% \\
     + LMFT (Joint Fine-tuning) & & & 0.22\% \\
     + Joint Fine-tuning & & & 0.28\% \\
     ++ LMFT & \multirow{-4}*{VoxCeleb2} & \multirow{-4}*{580+6.2M} & 0.18\% \\
     \midrule
     LoRA\_Adapter\_MFA(freeze PTM) & &  & 0.30\% \\
     + LMFT (Joint Fine-tuning) & & \multirow{-2}*{580+12.5M} & 0.22\% \\
     \cline{3-3}
     + Joint Fine-tuning (LoRA merge) & & & 0.30\% \\
     ++ LMFT & \multirow{-4}*{VoxCeleb2} & \multirow{-2}*{580+6.2M} & 0.24\% \\
    \midrule
     LoRA\_Adapter\_MFA(freeze PTM) & &  & 0.27\% \\
     + LMFT (Joint Fine-tuning) & & \multirow{-2}*{580+12.5M} & 0.15\% \\
     \cline{3-3}
     + Joint Fine-tuning (LoRA merge) & & & 0.23\%  \\
     ++ LMFT & \multirow{-4}*{\parbox{1cm}{\centering VoxCeleb2 \\ \& \\ VoxBlink2}} & \multirow{-2}*{580+6.2M} & 0.14\% \\
    \midrule
    & & & \textbf{CN-Celeb Test EER} \\
    \midrule
     Adapter\_MFA(freeze PTM) & &  & 6.51\% \\
     + Joint Fine-tuning & \multirow{-2}*{\parbox{1cm}{\centering CnCeleb1 \\ \&2}} & \multirow{-2}*{580+6.2M} & 5.17\% \\
    \midrule
     LoRA\_Adapter\_MFA(freeze PTM) & & 580+12.5M & 4.87\% \\
     + Joint Fine-tuning (LoRA merge) & \multirow{-2}*{\parbox{1cm}{\centering CnCeleb1 \\ \&2}} & 580+6.2M & 4.67\% \\
    \bottomrule
    \end{tabular}
\end{table}

Table \ref{tab:res1} shows the SV performance of different w2v-BERT 2.0 based model architectures, as described in Section \ref{sec:model_config}. The results show that the features extracted from the w2v-BERT 2.0  provide substantial benefits for the SV task.
Even with the simplest MFA structure, the model achieves an EER of 0.26\% on the Vox1-O test set. The introduction of the Layer Adapter effectively transforms the raw features from the PTM to better suit the SV task, while significantly reducing the number of parameters from 65.6M to 6.2M through dimensionality scaling, resulting in an improved EER of 0.18\%. For the model using the layer-wise weighted average method, although it is followed by a powerful ECAPA-TDNN network, the weighted summation of features across layers leads to greater information loss compared to feature concatenation, ultimately achieving an EER of only 0.22\%.
The use of the LoRA module has significantly enhanced training efficiency. In particular, during the PTM freezing phase, the model's performance on the Vox1-O test set improved from 0.43\% to 0.30\%, and on the CN-Celeb test set from 6.51\% to 4.87\%. Additionally, before fine-tuning with PTM unfrozen, the LoRA module's weights are merged into the PTM.
However, it is important to note that empolying the LoRA module on small and simple dateset may pose a risk of overfitting. For instance, when training only on the VoxCeleb2 dataset, performance slightly declined after unfreezing the PTM. After incorporating the VoxBlink2 dataset, the overfitting issue was effectively mitigated.

Table \ref{tab:res} reports a comparison between our w2v-BERT 2.0 based SV model and other SV models. After LMFT and score calibration, our model achieves an EER of 0.12\% on the Vox1-O test set, outperforming the SOTA ResNet293's result of 0.17\% EER~\cite{voxblink2}. 
Moreover, when trained only on the VoxCeleb2 dataset, the current SOTA PTM-based model yields 0.37\% EER on the Vox1-O test set~\cite{wavlm_lap_astp}, whereas our model achieves 0.18\% EER only using LMFT. Additionally, our model achieves an EER of 4.67\% on the CN-Celeb test set by only using the CnCeleb data for training, further demonstrating the robustness and generalization.



\begin{table}[t]\centering \scriptsize
    \caption{Results of knowledge distillation guided structured pruning on a w2v-BERT 2.0 based SV model trained on VoxCeleb2 and VoxBlink2. MACs and FLOPs were measured on 1-s long segments.}
    \tabcolsep=0.4em
     \label{tab:prune}
    \begin{tabular}{lcccccc}
    \toprule
    \textbf{Model}  & \textbf{Sparsity} & \textbf{Params} & \textbf{MACs} & \textbf{FLOPs} & \textbf{LMFT} &\textbf{Vox1-O EER}\\
    \midrule 
     \multirow{4}*{\shortstack[l]{LoRA\_Adapter\\\_MFA}} & \multirow{2}*{0\%} & \multirow{2}*{580+6.2M} & \multirow{2}*{28.75G} & \multirow{2}*{57.72G} & $\times$ & 0.23\% \\
      & & & & & $\checkmark$ & 0.14\% \\
     \cline{2-7}
      & \multirow{2}*{$\approx 80\%$} & \multirow{2}*{124+6.2M} & \multirow{2}*{6.31G} & \multirow{2}*{12.75G} & $\times$ & 0.35\% \\
      & & & & & $\checkmark$ & 0.18\% \\
    \bottomrule
    \end{tabular}
\end{table}




Table \ref{tab:prune} presents the results of knowledge distillation guided structured pruning applied to the w2v-BERT 2.0 based SV model. At a sparsity level of 80\%, our approach achieves an EER of 0.18\% on the Vox1-O test set after LMFT. Compared to the baseline system, the performance degradation is only 0.04\% EER, demonstrating promising potential for practical deployment.



\section{Conclusion}
In this paper, we explore the application of the w2v-BERT 2.0 PTM in the SV task. We adopt a Layer Adapter based MFA framework, combined with efficient fine-tuning via LoRA, to aggregate multi-layer features from the PTM and extract speaker embeddings. The experimental results show that our model achieves the SOTA performance, with an EER of 0.12\% on the Vox1-O test set and 4.67\% on the CN-Celeb test set. Furthermore, to enhance practical deployability, we apply knowledge distillation guided structured pruning, reducing the PTM’s parameter count by 80\% while incurring only a 0.04\% increase in EER. Source code and models are released.


\vfill\pagebreak

{
\small
\bibliographystyle{IEEEbib_new}
\bibliography{strings2}

\begin{thebibliography}{10}
\providecommand{\url}[1]{#1}
\csname url@samestyle\endcsname
\providecommand{\newblock}{\relax}
\providecommand{\bibinfo}[2]{#2}
\providecommand{\BIBentrySTDinterwordspacing}{\spaceskip=0pt\relax}
\providecommand{\BIBentryALTinterwordstretchfactor}{4}
\providecommand{\BIBentryALTinterwordspacing}{\spaceskip=\fontdimen2\font plus
\BIBentryALTinterwordstretchfactor\fontdimen3\font minus \fontdimen4\font\relax}
\providecommand{\BIBforeignlanguage}[2]{{%
\expandafter\ifx\csname l@#1\endcsname\relax
\typeout{** WARNING: IEEEtran.bst: No hyphenation pattern has been}%
\typeout{** loaded for the language `#1'. Using the pattern for}%
\typeout{** the default language instead.}%
\else
\language=\csname l@#1\endcsname
\fi
#2}}
\providecommand{\BIBdecl}{\relax}
\BIBdecl

\bibitem{voxceleb1}
A.~Nagrani, J.~S. Chung, and A.~Zisserman, ``Voxceleb: A large-scale speaker identification dataset,'' in \emph{Proc. Interspeech}, 2017, pp. 2616--2620.

\bibitem{vox2dev}
J.~Chung, A.~Nagrani, and A.~Zisserman, ``Voxceleb2: {Deep} {Speaker} {Recognition},'' in \emph{Proc. Interspeech}, 2018.

\bibitem{voxblink2}
Y.~Lin, M.~Cheng, F.~Zhang \emph{et~al.}, ``{VoxBlink2: A 100K+ Speaker Recognition Corpus and the Open-Set Speaker-Identification Benchmark},'' in \emph{{Proc. Interspeech}}, 2024, pp. 4263--4267.

\bibitem{cnceleb1}
Y.~Fan, J.~Kang, L.~Li \emph{et~al.}, ``Cn-celeb: a challenging chinese speaker recognition dataset,'' in \emph{Proc. ICASSP}, 2020, pp. 7604--7608.

\bibitem{cnceleb2}
L.~Li, R.~Liu, J.~Kang \emph{et~al.}, ``Cn-celeb: multi-genre speaker recognition,'' \emph{Speech Communication}, vol. 137, pp. 77--91, 2022.

\bibitem{ecapa-tdnn}
B.~Desplanques, J.~Thienpondt, and K.~Demuynck, ``Ecapa-tdnn: Emphasized channel attention, propagation and aggregation in tdnn based speaker verification,'' in \emph{Proc. Interspeech}, 2020, pp. 3830--3834.

\bibitem{cam++}
H.~Wang, S.~Zheng, Y.~Chen, L.~Cheng, and Q.~Chen, ``Cam++: A fast and efficient network for speaker verification using context-aware masking,'' in \emph{Proc. Interspeech}, 2023, pp. 5301--5305.

\bibitem{redimnet}
I.~Yakovlev, R.~Makarov, A.~Balykin \emph{et~al.}, ``{Reshape Dimensions Network for Speaker Recognition},'' in \emph{{Proc. Interspeech}}, 2024, pp. 3235--3239.

\bibitem{hubert}
W.-N. Hsu, B.~Bolte, Y.-H.~H. Tsai \emph{et~al.}, ``Hubert: Self-supervised speech representation learning by masked prediction of hidden units,'' \emph{IEEE/ACM Trans. ASLP}, vol.~29, pp. 3451--3460, 2021.

\bibitem{wav2vec2.0}
A.~Baevski, Y.~Zhou, A.~Mohamed, and M.~Auli, ``wav2vec 2.0: A framework for self-supervised learning of speech representations,'' in \emph{Proc. NeurIPS}, 2020, pp. 12\,449--12\,460.

\bibitem{unispeech}
S.~Chen, Y.~Wu, C.~Wang \emph{et~al.}, ``Unispeech-sat: Universal speech representation learning with speaker aware pre-training,'' in \emph{Proc. ICASSP}, 2022, pp. 6152--6156.

\bibitem{w2v-bert-2.0}
L.~Barrault, Y.-A. Chung, M.~C. Meglioli \emph{et~al.}, ``Seamless: Multilingual expressive and streaming speech translation,'' \emph{arXiv preprint arXiv:2312.05187}, 2023.

\bibitem{wavlm}
S.~Chen, C.~Wang, Z.~Chen \emph{et~al.}, ``Wavlm: Large-scale self-supervised pre-training for full stack speech processing,'' \emph{IEEE J-STSP}, vol.~16, no.~6, pp. 1505--1518, 2022.

\bibitem{hubert_w2v_unis_asv}
Z.~Chen, S.~Chen, Y.~Wu \emph{et~al.}, ``Large-scale self-supervised speech representation learning for automatic speaker verification,'' in \emph{Proc. ICASSP}, 2022, pp. 6147--6151.

\bibitem{wavlm_lap_astp}
J.~S. Kim, H.~J. Park, W.~Shin, and S.~W. Han, ``Rethinking leveraging pre-trained multi-layer representations for speaker verification,'' in \emph{Proc. Interspeech}, 2025, pp. 3713--3717.

\bibitem{wavlm_ca-mhfa}
J.~Peng, L.~Mo{\v{s}}ner, L.~Zhang, O.~Plchot, T.~Stafylakis, L.~Burget, and J.~{\v{C}}ernock{\`y}, ``Ca-mhfa: A context-aware multi-head factorized attentive pooling for ssl-based speaker verification,'' in \emph{Proc. ICASSP}, 2025, pp. 1--5.

\bibitem{whisper_asv}
Y.~Zhao, S.~Wang, G.~Sun, Z.~Chen, C.~Zhang, M.~Xu, and T.~F. Zheng, ``Whisper-pmfa: Partial multi-scale feature aggregation for speaker verification using whisper models,'' in \emph{Proc. Interspeech}, 2024, pp. 2680--2684.

\bibitem{cai_asr_sv}
D.~Cai and M.~Li, ``Leveraging asr pretrained conformers for speaker verification through transfer learning and knowledge distillation,'' \emph{IEEE/ACM Trans. ASLP}, vol.~32, pp. 3532--3545, 2024.

\bibitem{mfa}
Y.~Zhang, Z.~Lv, H.~Wu, S.~Zhang, P.~Hu, Z.~Wu \emph{et~al.}, ``{MFA-Conformer: Multi-scale Feature Aggregation Conformer for Automatic Speaker Verification},'' in \emph{Proc. Interspeech}, 2022, pp. 306--310.

\bibitem{lora}
E.~J. Hu, Y.~Shen, P.~Wallis, Z.~Allen-Zhu, Y.~Li, S.~Wang \emph{et~al.}, ``Lora: Low-rank adaptation of large language models.'' in \emph{Proc. ICLR}, 2022.

\bibitem{prune}
J.~Han, P.~P{\'a}lka, M.~Delcroix, F.~Landini, J.~Rohdin, J.~Cernock{\`y}, and L.~Burget, ``Efficient and generalizable speaker diarization via structured pruning of self-supervised models,'' \emph{arXiv preprint arXiv:2506.18623}, 2025.

\bibitem{w2v-bert}
Y.-A. Chung, Y.~Zhang, W.~Han \emph{et~al.}, ``W2v-bert: Combining contrastive learning and masked language modeling for self-supervised speech pre-training,'' in \emph{Proc. ASRU}, 2021, pp. 244--250.

\bibitem{asp_pooling}
K.~Okabe, T.~Koshinaka, and K.~Shinoda, ``Attentive statistics pooling for deep speaker embedding,'' in \emph{Proc. Interspeech}, 2018, pp. 2252--2256.

\bibitem{lglr}
Z.~Wang, J.~Wohlwend, and T.~Lei, ``Structured pruning of large language models,'' in \emph{Proc. EMNLP}, 2020, pp. 6151--6162.

\bibitem{3dspeaker}
Y.~Chen, S.~Zheng, H.~Wang, L.~Cheng, T.~Zhu, R.~Huang \emph{et~al.}, ``3d-speaker-toolkit: An open-source toolkit for multimodal speaker verification and diarization,'' in \emph{Proc. ICASSP}, 2025, pp. 1--5.

\bibitem{wespeaker}
H.~Wang, C.~Liang, S.~Wang, Z.~Chen, B.~Zhang, X.~Xiang \emph{et~al.}, ``Wespeaker: A research and production oriented speaker embedding learning toolkit,'' in \emph{Proc. ICASSP}, 2023, pp. 1--5.

\bibitem{on-the-fly}
W.~{Cai}, J.~{Chen}, J.~{Zhang}, and M.~{Li}, ``{On-the-Fly Data Loader and Utterance-Level Aggregation for Speaker and Language Recognition},'' \emph{IEEE/ACM Trans. ASLP}, pp. 1038--1051, 2020.

\bibitem{musan}
D.~Snyder, G.~Chen, and D.~Povey, ``Musan: A music, speech, and noise corpus,'' \emph{arXiv preprint arXiv:1510.08484}, 2015.

\bibitem{RIR}
T.~Ko, V.~Peddinti, D.~Povey, M.~Seltzer, and S.~Khudanpur, ``A study on data augmentation of reverberant speech for robust speech recognition,'' in \emph{Proc. ICASSP}, 2017, pp. 5220--5224.

\bibitem{speed_aug}
W.~Wang, D.~Cai, X.~Qin, and M.~Li, ``The dku-dukeece systems for voxceleb speaker recognition challenge 2020,'' \emph{arXiv preprint arXiv:2010.12731}, 2020.

\bibitem{adamw}
I.~Loshchilov and F.~Hutter, ``Decoupled weight decay regularization,'' in \emph{Proc. ICLR}, 2019.

\bibitem{arcface}
J.~Deng, J.~Guo, N.~Xue, and S.~Zafeiriou, ``Arcface: Additive angular margin loss for deep face recognition,'' in \emph{{Proc. CVPR}}, 2019, pp. 4690--4699.

\bibitem{lmft}
J.~Thienpondt, B.~Desplanques, and K.~Demuynck, ``The idlab voxsrc-20 submission: Large margin fine-tuning and quality-aware score calibration in dnn based speaker verification,'' in \emph{Proc. ICASSP}, 2021, pp. 5814--5818.

\bibitem{As-norm}
P.~Matejka, O.~Novotn{\`y}, O.~Plchot \emph{et~al.}, ``Analysis of score normalization in multilingual speaker recognition.'' in \emph{Proc. Interspeech}, 2017, pp. 1567--1571.

\bibitem{qmf}
Z.~Li, Y.~Lin, X.~Qin, N.~Jiang, G.~Zhao, and M.~Li, ``The dku-msxf speaker verification system for the voxceleb speaker recognition challenge 2023,'' \emph{arXiv preprint arXiv:2308.08766}, 2023.

\end{thebibliography}
}
\end{document}